\documentclass[letters]{mnras}

\usepackage{mathptmx}

\usepackage[T1]{fontenc}
\usepackage{ae,aecompl}

\usepackage{graphicx}   
\usepackage{amsmath}    
\usepackage{amssymb}    

\title[Estimation of the mass of Cygnus X-1]{ 
Constraining Mass of Cygnus X-1 from Analysis of the Hard State Spectral Data using TCAF Solution}

\author[Banerjee et al.]{
I. Banerjee$^{1}$\thanks{E-mail: tpib@iacs.res.in, (IACS)},
A. Bhattacharjee$^{2}$\thanks{E-mail: ayan12@bose.res.in, (SNBNCBS)},
A. Banerjee$^{2}$\thanks{E-mail: anuvab.banerjee@bose.res.in, (SNBNCBS)},
D. Debnath$^{3}$\thanks{E-mail: dipak@csp.res.in, (ICSP)}
and S. K. Chakrabarti$^{3}$\thanks{E-mail: sandip@csp.res.in, (ICSP)}
\\
$^{1}$Indian Association for the Cultivation of Science, 2A \& 2B Raja S C Mullick Road, Kolkata 700032, India\\
$^{2}$S. N. Bose National Centre for Basic Sciences, Salt Lake, Kolkata 700106, India\\
$^{3}$Indian Center for Space Physics, 43 Chalantika, Garia St. Road, Kolkata 700084, India\\
}

\begin{document}
\label{firstpage}
\pagerange{\pageref{firstpage}--\pageref{lastpage}}
\maketitle

\begin{abstract}
The  galactic black hole candidate Cygnus X-1, one of the brightest sources in the sky, is the first ever black hole candidate to be discovered. Despite being a very well-studied object due to its persistent brightness in X-rays, there has been much difficulty in determining its mass since its discovery. In spite of continuous efforts since the early nineteen seventies, there is yet no concensus on its mass for nearly four decades. The uncertainties in the mass measurements are due to the high degree of error  involved in the measurement of its distance. In 2011, Orosz et al. constrained the mass of this object to be $M= 14.8 \pm 1.0 M_\odot$ using dynamical methods. In this paper, we obtained the mass of Cygnus X-1, using a completely independent method, namely,
carrying out the spectral analysis using Two Component Advective Flow (TCAF) solution based {\it fits} file and the archival data of RXTE PCA instrument. Our result does not require the distance of the source os the information about the companion. Each spectral fit with the TCAF gives one best fitted mass. Averaging fitted masses of Cygnus X-1 over a span of five months of observation during its persistent hard phase, mass of the source comes out to be $M_{avg}=14.20 \pm 0.36 M_\odot$, which is consistent with the dynamically estimated mass.

\end{abstract}

\begin{keywords}
X-Rays:binaries - stars individual: (Cygnus X-1) - stars:black holes - accretion, accretion disks - shock waves - radiation:dynamics
\end{keywords}




\section{Introduction}
Cygnus X-1, the first ever stellar-mass black hole candidate (BHC) to be discovered is a very important source to study physics of accretion processes around black holes due to its proximity and persistent brightness in X-rays.
Over the past several decades, various methods have been employed for the determination of its mass and different estimates have been reported in the literature. However, most of these estimates rely on the mass of the companion or the distance of the compact object. Based on the radial velocity profile of its companion, Bolton (1972) and Webster \& Murdin (1972) inferred that the compact object in Cygnus X-1 should be a black hole. Bolton (1972) gave lower limits to the the mass of the black hole and its companion to be $3 M_{\odot}$ and $12 M_{\odot}$ respectively while Paczy\'{n}ski (1974) estimated the lower limit of the black hole mass to $3.6 M_{\odot}$ assuming the distance $D > 1.4 \rm kpc$.
Gies \& Bolton (1986) subsequently reported that $M_{BH}> 7 M_\odot$ assuming the mass of the secondary to
be $M \sim 16 M_\odot$. Considering the companion mass of $\sim 20 M_\odot$, Ninkov et al. (1987) estimated the mass of the compact object to be $10 \pm 1 M_{\odot}$ while Herrero et al. (1995) reported the mass of the black hole to be $4.8 M_{\odot} < M _{BH}< 14.7 M_{\odot}$. All these mass measurements are subject to several uncertainties because of the high degree of errors associated with the estimation of the distance and companion properties which in turn affect  the model parameters on which the mass depends. 
Clearly, the measurement of distance is of primary importance in order to obtain an accurate 
constraint on the mass of the black hole if one is limited to dynamical methods. Reid et al. (2011) estimated the distance of the source to be $1.86^{+0.12}_{-0.11}$ kpc, based on the trigonometric parallax method. Knowing the distance and the orbital velocity of the companion, Orosz et al. (2011) estimated the  
inclination angle of the source to be $\theta = 27.1^o \pm 0.8^o$ which subsequently enabled them to provide a stronger constraint on the mass of the BHC in Cygnus X-1 to be $M_{BH}=14.8 \pm 1.0$ $M_{\odot}$.
Condidering the controversies associated with the mass of Cygnus X-1, it is important to estimate the mass of this object by other independent methods which do not depend on the distance or the mass of the companion. The aim of this paper is therefore to provide a constraint on the mass of Cygnus X-1 from the spectral analyses using the TCAF solution.   

In TCAF solution (Chakrabarti 1989; 1995, 1997) the accretion flow is comprises of two dynamical components,  namely, the viscous Keplerian flow along the equatorial plane and the low-angular momentum, weakly viscous, rapidly falling, sub-Keplerian flow enveloping the Keplerian disk. The advective nature of the sub-Keplerian flow enables it to attain supersonic speed far away from the black hole, which however brakes due to rapid strengthening of the centrifugal barrier (centrifugal force $\sim 1/r^3$ as opposed to gravity $\sim 1/r^2$) by means of a shock transition. The post-shock flow gets distended due to shock heating giving rise to the CENtrifugal barrier supported BOundary Layer (CENBOL) which acts as the `Compton Cloud'.  While analyzing RXTE data, several observers commented that such components have indeed been observed (e.g., Smith et al., 2002; Wu et al., 2002; Cambier \& Smith, 2013; Tomsick et al., 2014b). 

TCAF spectral model has been successfully implemented within XSPEC as a local additive table model (Debnath et al. 2014, 2015a,b). Estimates of the mass of several black hole sources (Molla et al. 2016, 2017; Jana et al. 2016,; Chatterjee et al. 2016, 2019; Debnath et al., 2017; Bhattacharjee et al. 2017; Shang et al. 2019) have been made using TCAF. For Cygnus X-1, TCAF was used by Ghosh et al. (2019) to determine the size of disc, through spectral analysis. Here, we focus only on the determination of mass from each spectral fit and estimate the average mass. 
To achieve a fit using TCAF, one requires four flow paramaters: 1. the Keplerian/disk rate ($\dot{m}_d$), 2. sub-Keplerian/halo rate ($\dot{m}_h$), 3. the location of the shock ($X_s$) and 4. the strength of the shock ($R$) (Chakrabarti 1995, for several spectra, see, Chakrabarti, 1997). There are two more parameters intrinsic to the source: 1. the mass of the black hole $M_{BH}$ and 2. a suitable constant normalization $N$ which is required to match the observed spectra with the emitted spectra across
the spectral states.
The combined six parameters are used here to fit each spectra. It is worth noting that the TCAF paradigm can also be applied to the cases of weakly magnetic neutron stars, with certain modifications (Bhattacharjee and Chakrabarti, 2017; Bhattacharjee 2018, Bhattacharjee \& Chakrabarti 2019).

We have analyzed the spectral data of Cygnus X-1 with the TCAF solution based additive table {\it fits} file to determine the mass. For this purpose we pick 20 observations from a hard state where the other TCAF parameters, related to the flow dynamics are almost constant. In this \textit{letter} we report the cases corresponding to the data obtained bewteen December 10, 1997 to May 12, 1998. 
In \S 2 we explain the observation and the data analysis procedure.
In \S 3 we present the results obtained from the 
spectral analysis. Finally in \S 4, we conclude with a brief summary of the results obtained.

\section{Observation and Data Analysis}
Long-term spectral analysis of Cygnus X-1 reveal that the object mainly exhibits two types of spectral states, either a prolonged  hard state or an extended period of soft state (Zhang et al. 1997, Cui et al. 1997, 1998 and references therein). 
From 1996 (RXTE launch) to 2015, the states of Cygnus X-1 in different major observation periods have been vividly reported by Grinberg et al. 2013. We report results of our analysis of the RXTE/PCA archival data of Cygnus X-1 from December 10, 1997 (MJD=50792.297) to May 12, 1998 (MJD=50945.736) when the object resided in the stable hard state. This is because, the hard phase is dominated by a prominent and widely extended CENBOL region which coincides with the inner edge of a Keplerian disk. Consequently, we expect that the effect of the black hole spin will not affect the estimates of the flow parameters or mass during the low/hard phase. 

For data reduction and spectral fitting we follow the method as described in Debnath et al. (2013, 2015a). 
The 2.5 to 45.0 keV background subtracted PCA spectra are fitted with the TCAF solution based {\it fits} file in XSPEC 
as an additive table model. The systematic error of 1\% has been employed. The hydrogen column density $N_H$ for absorption 
model {\it phabs} is kept roughly in the range $(0.48 - 10.0) \times 10^{22}$~atoms~cm$^{-2}$ (Grinberg et al. 2015), 
in order to obtain the best fits.

\setlength{\tabcolsep}{0.3em}
\begin{table*}
\small
\caption{Variation of the TCAF fit parameters with MJD keeping all the parameters free, in the 2.5-45.0 keV
energy range. Duration of the data is from December 10, 1997 to May 12, 1998.}
\begin{center}
\begin{tabular}{c c c c c c c c c c c}
\hline
\hline
$Obs$ & $\rm Id$ & $\rm MJD$ & $ N_H $ & $\rm \dot{m}_d ~ (\dot{M}_{Edd})$ & $\rm \dot{m}_h ~(\dot{M}_{Edd})$ & $M_{BH} ~ (M_{\odot})$ & $X_s ~ (r_S)$ & $ R $ & $N$ & $\chi^2/dof$ \\
\hline

$1$&$30158-01-01-00$ & $50792.297$ & $ 0.462 $ &    $0.711^{+0.009}_{-0.011}$ & $1.801^{+0.001}_{-0.001}$ & $14.623^{+0.025}_{-0.587}$& $72.818^{+0.615}_{-2.593}$ & $1.238^{+0.001}_{-0.002}$ & $2.433^{+1.031}_{-0.031}$ & $90.25/70$ \\

$2$&$30158-01-03-00$ & $ 50796.368$ & $ 0.517 $ & $0.709^{+0.011}_{-0.013}$ & $ 1.824^{+0.001}_{-0.001}$ & $ 14.493^{+0.084}_{-0.088}$ & $ 73.158^{+0.735}_{-2.332}$ & $ 1.225^{+0.003}_{-0.001}$ & $ 2.592^{+0.035}_{-0.034}$ & $ 79.32/70$\\

$3$&$30158-01-06-00$ & $50799.028$&$ 0.430  $ & $  0.730^{+0.003}_{-0.003}$&$ 1.735^{+0.003}_{-0.002}$&$ 13.880^{+0.119}_{-0.218}$&$ 72.315^{+5.073}_{-0.667}$&$ 1.186^{+0.001}_{-0.001}$&$ 2.180^{+0.021}_{-0.022}$&$ 93.40/70$\\

$4$&$30158-01-08-00$ & $ 50803.229$&$ 0.552 $ & $  0.720^{+0.003}_{-0.010}$&$ 1.833^{+0.001}_{-0.002}$&$ 13.979^{+1.01}_{-0.299}$&$ 71.999^{+2.014}_{-0.986}$&$ 1.217^{+0.004}_{-0.001}$&$ 2.803^{+0.018}_{-0.028}$&$ 108.36/70$\\

$5$&$30158-01-10-00$ & $ 50807.027$ & $ 0.896 $ & $0.716^{+0.006}_{-0.011}$ &$ 1.810^{+0.001}_{-0.002}$ &$ 14.006^{+0.084}_{-0.584}$ &$ 72.420^{+0.768}_{-2.643}$ &$ 1.241^{+0.002}_{-0.002}$ &$ 2.150^{+0.029}_{-0.031}$ &$ 87.35/70$\\

$6$&$30161-01-01-00$ & $ 50810.879 $ &$ 0.780 $ & $ 0.711^{+0.009}_{-0.003}$ &$ 1.722^{+0.012}_{-0.016}$ &$ 13.597^{+0.037}_{-0.313}$ &$ 72.509^{+3.927}_{-1.067}$ &$ 1.180^{+0.001}_{-0.001}$ &$ 2.147^{+0.021}_{-0.025}$ &$ 93.09/70$\\

$7$&$30157-01-04-00$ & $ 50812.763$ &$ 0.758  $ & $ 0.716^{+0.006}_{-0.012} $ &$1.824^{+0.001}_{-0.001}$ &$ 14.290^{+0.057}_{-0.436}$ &$ 72.396^{+1.031}_{-1.450}$ &$1.227^{+0.002}_{-0.002}$ &$ 2.171^{+0.030}_{-0.031}$ &$ 75.29/70$\\

$8$&$20175-01-03-00$ & $ 50815.031$ &$ 0.511  $ & $ 0.713^{+0.008}_{-0.007}$ &$ 1.733^{+0.002}_{-0.002}$ &$ 14.323^{+0.026}_{-0.261}$ &$ 72.909^{+2.282}_{-1.120}$ &$ 1.183^{+0.001}_{-0.001}$ &$ 1.862^{+0.021}_{-0.021}$ &$ 51.76/70$\\

$9$&$30157-01-05-00$ & $ 50821.198$ &$  0.518  $ & $  0.716^{+0.007}_{-0.018}$ &$ 1.819^{+0.001}_{-0.003}$ &$ 14.174^{+0.035}_{-0.075}$ &$ 72.766^{+1.066}_{-2.026}$ &$ 1.229^{+0.003}_{-0.002}$ &$ 2.206^{+0.047}_{-0.039}$ &$ 65.17/70$\\

$10$&$30157-01-08-00$ & $ 50843.049$ &$ 0.540  $ & $ 0.719^{+0.003}_{-0.013}$ &$ 1.827^{+0.001}_{-0.002}$ &$ 14.077^{+0.067}_{-0.472}$ &$ 72.071^{+1.401}_{-1.209}$ &$ 1.224^{+0.002}_{-0.002}$ &$ 3.034^{+0.039}_{-0.043}$ &$ 118.01/70$\\

$11$&$30157-01-09-01$ & $ 50849.981 $ &$ 0.765 $ & $ 0.723^{+0.003}_{-0.003}$ &$ 1.828^{+0.002}_{-0.003}$ &$ 14.086^{+0.310}_{-0.333}$ &$ 72.330^{+1.244}_{-1.072}$ &$ 1.224^{+0.002}_{-0.002}$ &$ 2.661^{+0.009}_{-0.038}$ &$92.04/70$\\

$12$&$30157-01-12-00$ & $ 50872.114 $ &$ 0.655  $ & $0.723^{+0.005}_{-0.001}$ &$ 1.839^{+0.001}_{-0.003}$ &$ 13.945^{+0.139}_{-0.390}$ &$ 72.100^{+0.395}_{-1.162}$ &$ 1.218^{+0.002}_{-0.001}$ &$ 3.313^{+0.008}_{-0.025}$ &$ 109.70/70$\\

$13$&$30157-01-13-00$ & $ 50882.784$ &$ 0.582 $ &  $  0.716^{+0.006}_{-0.010}$ &$ 1.807^{+0.001}_{-0.002}$ &$ 14.487^{+0.268}_{-0.670}$ &$ 71.506^{+1.679}_{-2.834}$ &$ 1.241^{+0.002}_{-0.002}$ &$ 2.093^{+0.028}_{-0.028}$ &$ 85.20/70$\\

$14$&$30157-01-14-00$&$ 50888.785$ &$ 0.781  $ & $ 0.704^{+0.012}_{-0.012}$ &$ 1.794^{+0.001}_{-0.001}$ &$ 13.634^{+0.074}_{-0.076}$ &$ 73.350^{+0.490}_{-0.502}$ &$ 1.251^{+0.002}_{-0.002}$ &$ 2.500^{+0.035}_{-0.034}$ &$ 94.36/70$\\

$15$&$30157-01-17-00$&$ 50908.862 $ &$ 0.611  $ & $0.723^{+0.004}_{-0.002}$ &$ 1.825^{+0.002}_{-0.002}$ &$ 14.679^{+0.112}_{-0.442}$ &$ 72.264^{+1.330}_{-1.288}$ &$ 1.224^{+0.002}_{-0.001}$ &$ 2.346^{+0.028}_{-0.031}$ &$ 92.48/70$\\

$16$&$30157-01-18-00$&$ 50916.137$ &$ 0.641  $ & $ 0.724^{+0.003}_{-0.003}$ &$ 1.818^{+0.002}_{-0.002}$ &$ 14.702^{+0.232}_{-0.390}$ &$ 71.938^{+1.868}_{-1.170}$ &$ 1.226^{+0.002}_{-0.002}$ &$ 2.692^{+0.037}_{-0.034}$ &$ 72.09/70$\\

$17$&$30157-01-19-00$&$ 50924.929 $ &$ 0.488 $ & $ 0.724^{+0.004}_{-0.004}$ &$ 1.803^{+0.003}_{-0.002}$ &$ 13.671^{+0.490}_{-0.514}$ &$ 70.379^{+2.830}_{-2.344}$ &$ 1.242^{+0.003}_{-0.002}$ &$ 2.671^{+0.036}_{-0.038}$ &$ 105.15/70$\\

$18$&$30157-01-20-00$&$ 50931.595$ &$ 0.487  $ & $ 0.716^{+0.006}_{-0.011}$ &$ 1.800^{+0.001}_{-0.002}$ &$ 14.513^{+0.138}_{-0.647}$ &$ 71.994^{+1.170}_{-2.866}$ &$ 1.244^{+0.002}_{-0.002}$ &$ 2.096^{+0.028}_{-0.028}$ &$ 90.17/70$\\

$19$&$30157-01-21-00$&$ 50939.009$ &$ 0.599  $ & $ 0.724^{+0.004}_{-0.002}$ &$ 1.815^{+0.002}_{-0.002}$ &$ 14.152^{+0.175}_{-0.383}$ &$ 72.068^{+1.985}_{-1.181}$ &$ 1.225^{+0.002}_{-0.002}$ &$ 2.921^{+0.043}_{-0.031}$ &$ 114.71/70$\\

$20$&$30157-01-22-00$ & $ 50945.736$ &$ 1.143  $ & $ 0.679^{+0.010}_{-0.010}$ &$ 1.810^{+0.001}_{-0.001}$ &$ 14.817^{+0.081}_{-0.484}$ &$ 72.303^{+0.929}_{-1.519}$ &$ 1.229^{+0.001}_{-0.002}$ &$ 2.188^{+0.027}_{-0.028}$ &$ 95.22/70$\\

\hline

\end{tabular}
\label{tcaf_fits:nfree}
\end{center}
\end{table*}

\begin{figure}
  \centering
\includegraphics[height=8.0cm,width=8.0cm]{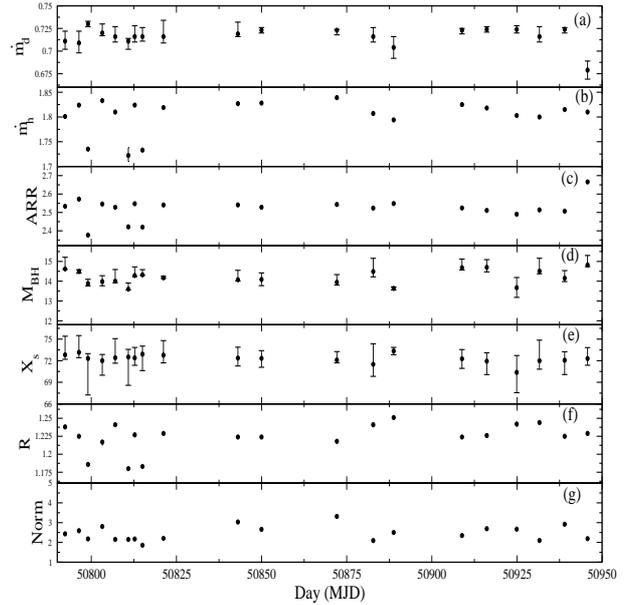}
\caption{Variation of
(a) the disk accretion rate $\dot{m}_d$ (in Eddington units),
(b) the sub-Keplerian halo accretion rate  $\dot{m}_h$ (in Eddington units),
(c) the accretion rate ratio (ARR) $\dot{m}_h/\dot{m}_d$,
(d) the mass of the black hole (in units of $M_{\odot}$),
(e) the shock location $X_s$ (in units of $r_S=2GM/c^2$),
(f) the shock strength $R$, and (g) the normalization of the TCAF model, with day (MJD).
Variation of all the aforementioned quantities are studied in the 2.5-45.0 keV energy band.
}
\end{figure}

\begin{figure}
  \centering
\includegraphics[height=8.0cm,width=8.0cm,angle=270]{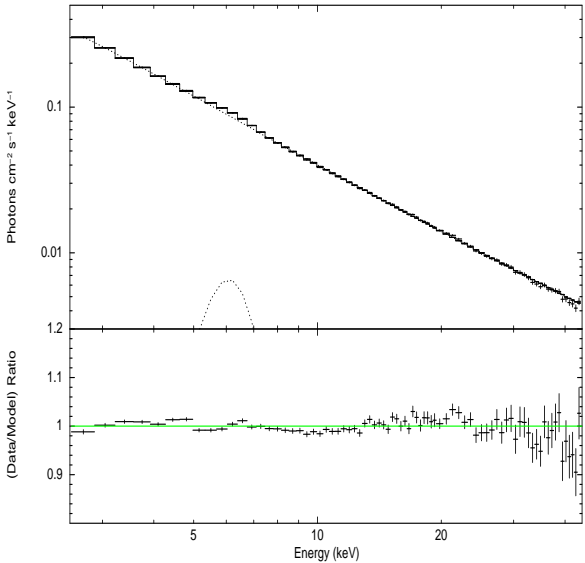}
\caption{Upper Panel: Unfolded spectra of observation ID 30157-01-18-00 for energy 2.5-45.0 keV,
fitted with \textit{phabs*(TCAF+Gaussian)} models showing the spectrum of a typical hard state. Lower Panel: The data/model ratio of the same spectral fit.}
\end{figure}

\section{Spectral data fitted by TCAF model during hard phase}
The spectra in the 2.5-45.0 keV energy range have been fitted using \textit{TCAF+Gaussian} model. The Gaussian component was incorporated to ensure the best fit (representing the contribution from Iron emission line) with peak energy between $6.2$ to $6.8$ keV. The width of the Gaussian component was allowed to vary in the range $0.001 - 0.8$ keV. The normalization of the Gaussian component was allowed to be
free, however. It was found that in order to obtain the best
fit, the normalization has to be varied between $7.25 \times 10^{-3}$
to $1.59 \times 10^{-2}$ during the hard-phase. The actual values of the fitted parameters are listed in Table 1. 

The disk accretion rate $\dot{m}_d$ is found to vary between $0.68 \dot{M}_{Edd}$ to $0.73 \dot{M}_{Edd}$ during this time period with an average of $\sim 0.72 \dot{M}_{Edd}$. The halo accretion rate $\dot{m}_h$, varies from $1.72 \dot{M}_{Edd}$ to $1.84 \dot{M}_{Edd}$ with an average of $\sim 1.80 \dot{M}_{Edd}$. Since the variation in both the accretion rates is minimal, the accretion rate ratio ARR remains roughly constant with slight fluctuations (see, Fig. 1(c)).
The average location of the shock is found to be $X_s \sim 72.28 r_S$, with minor variation between $70.38 r_S$ to $73.35 r_S$, where $r_S=2GM/c^2$ is the Schwarzschild radius. The average strength of the shock is obtained as $R \sim 1.22$, varying between $R=1.18$ to $R=1.25$. It is apparent that all the flow parameters vary within a very narrow range which is characteristic of the persistent hard phase. This is illustrated in Fig. 1. (a-c), (e-f)). The spectrum of a typical hard-state (MJD=50916.137) is illustrated in Fig. 2. 

Since the mass of the black hole $M_{BH}$ is a parameter of the TCAF, we obtained a constraint on the mass of the source from each of the spectral fits. The average mass obtained from our fits of the hard phase is $M_{avg}=14.20 \pm 0.36 M_\odot$. This is  in agreement with the estimates reported by some previous authors.  

\section{Discussions}
Cygnus X-1 is a well studied black hole candidate, and yet, the determined mass had a large range of uncertainty because of 
its dependence on the distance and the companion. It was therefore essential to have a method which is independent of these poorly known parameters.
In this letter, we used TCAF solution to successfully fit the spectra of Cygnus X-1 over a period of 
five months during one of its low/hard phase (between the end of 1996 to early 1998, Grinberg et al. 2013). This enables us to give an estimate of the underlying accretion flow parameters during this period. The behaviour of flow parameters, $\dot{m}_d$, $\dot{m}_h$, $X_s$ and $R$, was consistent with the hard state. 
It was shown by Grinberg et al. (2013) that no disc component was required for spectral fitting during the hard state, 
suggesting that the disc resided away from the source and it was small in size. This was further proven with
TCAF by Ghosh \& Chakrabarti (2016), Ghosh et al. (2018) and Ghosh et al. (2019). We also found the average location of the shock $X_s \sim 72.28 r_S$ indicating that the Keplerian disk is away from the black hole during the period of observation. This in turn removed the possibility of having any observable 
contribution from the Comptonized component of the iron line (reflection component), 
unlike the ones reported during the soft state (Tomsick et al. 2014a). 
This was also evident from the data/model ratio plot which remained almost constant around the value 1 till 45 keV (Figure 2). 
From each spectral fit we determine the mass of the BHC which lies well within the range of mass of the source 
estimated earlier by Orosz et al. (2011). The mass of Cygnus X-1 averaged over all the reported observations is found to be $M_{BH}=14.20 \pm 0.36 M_\odot$ which gives a much tighter constraint compared to the previous estimates reported in the literature. Furthermore, our estimated value is independent of any other 
parameters, such as the distance or the properties of the companion.


\begin{thebibliography}{50}
\bibitem{nsreview}Bhattacharjee, A., 2018, in Exploring the Universe: From Near Space to Extra-galactic Mukhopadhyay, B \& Sasmal, S. (Eds.), ASSP, 53, 93, Springer (Heidelberg)
\bibitem{}Bhattacharjee, A. et al. 2017, MNRAS, 466, 1372
\bibitem{2017MNRAS.472.1361B} Bhattacharjee, A., \& Chakrabarti, S.~K., 2017, MNRAS, 472, 1361
\bibitem{abskc2019}Bhattacharjee, A., \& Chakrabarti, S.~K., 2019, ApJ, 873, 119
\bibitem{}Bolton, C. T., 1972, Nature, 235, 271
\bibitem{}Cambier, H. J., \& Smith, D. M., 2013, ApJ, 767, 46
\bibitem{}Chakrabarti S. K., 1989, MNRAS, 240, 7
\bibitem{}Chakrabarti S. K., 1995, in Bohringer H., Morfil G. E., Trumper J., eds, Seventeenth Texas Symposium on Relativistic Astrophysics and Cosmology, Vol. 759. New York Academy of Sciences, New York, p.  546
\bibitem{}Chakrabarti S. K., 1997, ApJ, 484, 313
\bibitem{}Chatterjee, D. et al. 2016, ApJ, 827, 88
\bibitem{}Chatterjee, D., Debnath, D., Jana, A., et al., 2019, Ap\&SS, 364, 14
\bibitem{}Cui, W. et al. 1997, ApJ, 474, L57
\bibitem{}Cui, W. et al. 1998, ApJ, 493, L75
\bibitem{}Debnath, D., Chakrabarti, S.K. \& Mondal, S., 2014, MNRAS, 440, L121 
\bibitem{}Debnath, D., Mondal, S., \& Chakrabarti, S.K., 2015a, MNRAS, 447, 1984 
\bibitem{}Debnath, D. et al. 2015b, ApJ, 803, 59
\bibitem{}Debnath, D. et al. 2017, ApJ, 850, 92
\bibitem{}Dutta, B. G. \& Chakrabarti, S. K., 2010, MNRAS, 404, 2136
\bibitem{}Ghosh, A., \& Chakrabarti, S. K., 2016, Ap\&SS, 361, 310
\bibitem{}Ghosh, A., \& Chakrabarti, S. K., 2018, MNRAS, 479, 1210
\bibitem{}Ghosh, A., Banerjee, I., \& Chakrabarti, S. K., 2019, 484, 5802
\bibitem{}Gies, D. R., \& Bolton, C. T. 1986, ApJ, 304, 371
\bibitem{}Gleissner, T. et al., 2004, A \& A, 425, 1061
\bibitem{}Grinberg, V. et al., 2013, A\& A, 554, A88
\bibitem{}Grinberg, V. et al., 2015, A\&A, 576, A117
\bibitem{}Herrero, J. et al. 1995, A\& A, 297, 556
\bibitem{}Jana, A. et al., 2016, ApJ, 819, 107
\bibitem{}Molla, A. A. et al. 2017, ApJ, 834, 88
\bibitem{}Molla, A. A., Debnath, D., Chakrabarti, S. K., \& Mondal, S., Jana, A. 2016, MNRAS, 460, 3163
\bibitem{}Ninkov, Z., Walker, G. A. H., \& Yang, S. 1987, ApJ, 321, 425
\bibitem{}Orosz, J. A. et al., 2011, ApJ, 742, 84
\bibitem{}Paczy{\'{n}}ski, B. 1974, A\&A, 34, 161
\bibitem{}Reid, M. J. et al. 2011, ApJ, 742, 83
\bibitem{}Shang, J.R., Debnath, D., Chatterjee, D., et al., 2019, ApJ (in press)
\bibitem{}Shaposhnikov, N., \& Titarchuk, L. 2007, ApJ, 663, 445
\bibitem{}Smith, D., Heindl, W.A., \& Swank, J.H., 2002, ApJ, 569, 362
\bibitem{}Tomsick, J. A. et al., 2014a, ApJ, 780, 78
\bibitem{}Tomsick, J. A. et al., 2014b, ApJ, 791, 70
\bibitem{}Webster, B. L., \& Murdin, P. 1972, Nature, 235, 37
\bibitem{}Wu, K. et al., 2002, ApJ, 565, 1161
\bibitem{}Zhang, S. N., Cui, W., Harmon, B. A. et al. 1997, ApJ, 477, L95
\end{thebibliography}
\end{document}